\documentclass[11pt,twoside]{article}
\usepackage{asp2010} 
\resetcounters 

\aspvolume{455}
\aspcpryear{2012}
\aspvoltitle{4$^{th}$ {\em Hinode} Science Meeting: Unsolved 
Problems and Recent Insights}
\aspvolauthor{L.~R.~Bellot Rubio, F.~Reale, and M.~Carlsson, eds.}
\usepackage{natbib}

\begin{document} 
 
\title{Observations of Supersonic Downflows near the Umbra-Penumbra Boundary of Sunspots as Revealed by {\em Hinode}} 
\author{Rohan~E.~Louis,$^1$ Luis~R.~Bellot Rubio,$^2$ Shibu~K.~Mathew,$^1$ and P.~Venkatakrishnan$^1$ 
\affil{$^1$Udaipur Solar Observatory, Physical Research Laboratory 
                     Dewali, Badi Road, Udaipur, 
	       	     Rajasthan - 313004, India} 
\affil{$^2$Instituto de Astrof\'{\i}sica de Andaluc\'{\i}a (CSIC), 
                     Apartado de Correos 3004, 
                     18080 Granada, Spain}}

\begin{abstract} High resolution spectropolarimetric observations by 
{\em Hinode} have revealed the existence of supersonic downflows at
the umbra-penumbra boundary of 3 sunspots \citep{2011ApJ...727...49L}.  These
downflows are observed to be co-spatial with bright penumbral
filaments and occupy an area greater than 1.6 arcsec$^2$. They are
located at the center-side penumbra and have the same polarity as the
sunspot which suggests that they are not associated with the Evershed
flow. In this paper we describe the supersonic velocities observed in
NOAA AR 10923 and discuss the photospheric as well as chromospheric
brightenings that lie close to the downflowing areas. Our observations
suggest that this phenomenon is driven by dynamic and energetic
physical processes in the inner penumbra which affect the chromosphere, 
providing new constraints to numerical models of sunspots. 
\end{abstract} 
 
\section{Introduction} 
\label{intro} 
The Evershed flow \citep[EF;][]{1909MNRAS..69..454E} is a distinct property
of sunspot penumbrae which exemplifies their filamentary structure
\citep[][and references therein]{2003A&ARv..11..153S}.  In the inner
penumbra the EF starts as upflows \citep{2006A&A...453.1117B, 2006ApJ...646..593R,
2009A&A...508.1453F} that turns into downflows in the mid and outer penumbra
\citep{1997Natur.389...47W, 1999A&A...349L..37S, 2003A&A...410..695M, 2004A&A...427..319B}.
 
In addition to the EF, other types of mass motions exist in the
penumbra, as reported recently by \citet{2010A&A...524A..20K} using {\em
Hinode} observations.  They detected small-scale downflowing patches
with velocities of $\sim$1~km~s$^{-1}$ which have the same polarity
as the parent sunspot.  Some of them also appear to be co-spatial with
chromospheric brightenings.  Based on their physical properties,
\citet{2010A&A...524A..20K} inferred that these weak downflows are different
from the Evershed flow returning to the photosphere, which sometimes
happens well within the penumbra
\citep{2004A&A...427..319B,2007ApJ...668L..91B,2008A&A...481L..21S}. 
 
\begin{figure}[t] 
\centering
\includegraphics[width=11cm,angle=90,bb=-7 100 595 792]{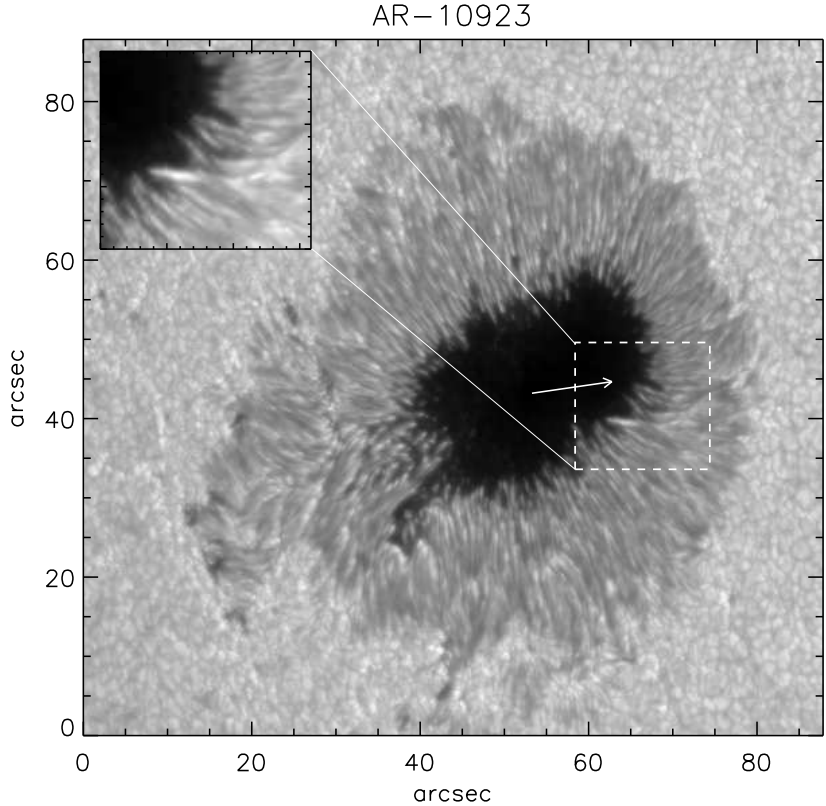}  
\caption{Continuum image of NOAA AR 10923 at 630~nm. The white dashed 
square represents the small region chosen for analysis and whose
magnified image is shown in the inset. The white arrow points to disk
center.}
\label{cont_image} 
\end{figure} 

\citet{2011ApJ...727...49L} observed a new type of downflows which are supersonic and 
occur at or near the umbra-penumbra boundary of sunspots. These
downflowing patches are conspicuously large in size with areas ranging
from 1.6--6 arcsec$^2$.  They have the same polarity as the sunspot and
occur along bright penumbral filaments. Their properties are different
from those of the Evershed flow and the downflows reported by
\citet{2010A&A...524A..20K}, which indicates a different physical
origin. The strong downflows possibly represent energetic and dynamic
processes occurring in the inner penumbra which also affect the
chromosphere.  In this paper we restrict our discussion to the
supersonic downflows observed in NOAA AR 10923 and briefly describe
the photospheric and chromospheric activities associated with them.

\section{Observations} 
\label{data} 
High resolution spectro-polarimetric observations of NOAA AR 10923
were carried out using the Solar Optical Telescope
\citep[SOT;][]{2008SoPh..249..167T} on board {\em Hinode} \citep{2007SoPh..243....3K}
on November 10, 2006 when the sunspot was located at a heliocentric
angle of $50\deg$ (Figure~\ref{cont_image}). The AR was
mapped by the {\em Hinode} spectro-polarimeter
\citep[SP;][]{2001ASPC..236...33L,2008SoPh..249..233I} from 16:01 to 17:25~UT in the  
normal map mode with an exposure time of 4.8~s and a pixel size of 0\farcs16. 
The four Stokes profiles of the neutral iron lines at 630~nm were recorded  
with a spectral sampling of 21.55~m\AA\/ at each slit position. In addition to the 
Stokes spectra, G-band and \ion{Ca}{ii} H filtergrams acquired by he Broadband Filter 
Imager (BFI) close to the SP scans were also employed. The filtergrams had a  
sampling of 0\farcs055 with a cadence of 30~s. These data were 
recorded from 13:00 to 14:00~UT. 
 
\begin{figure}[t] 
\centering
\includegraphics[width=6.5cm,angle=90,bb=100 00 526 842]{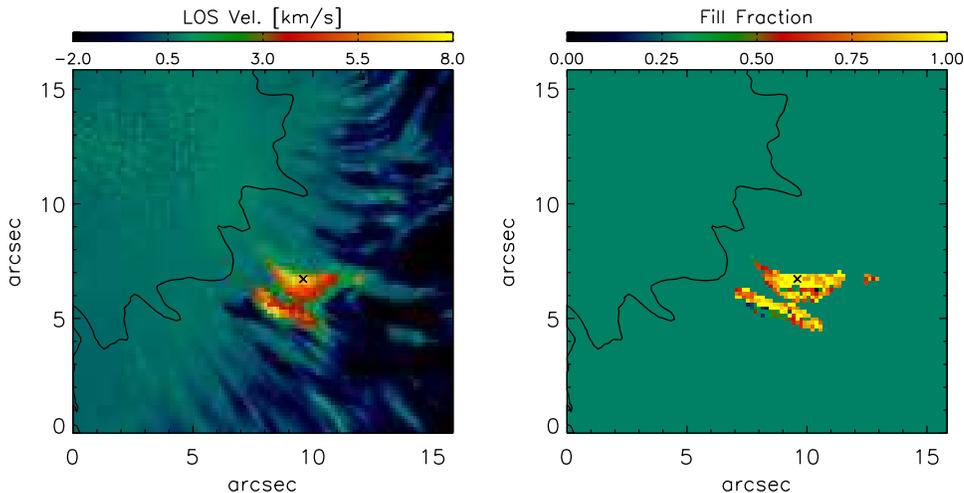} 
\caption{{\em Left:} LOS velocity derived from SIR combining one- and 
two-component inversions. {\em Right:} The fill fraction of the fast
component. The black contour depicts the umbra-penumbra boundary. Both
images have been scaled as shown in their respective color bars. The
black cross corresponds to the pixel exhibiting supersonic
velocity. The Stokes profiles emerging from this pixel are shown in
Figure~\ref{stokes}.}
\label{velo} 
\end{figure} 
 
\section{Results} 
\label{res} 
 
\subsection{Supersonic Downflows from SIR Inversions} 
\label{downflows} 
The detection of supersonic downflows was carried out by constructing
red and blue wing magnetograms at $\pm$34.4 pm from the line center of
the \ion{Fe}{i} 630.25~nm line as described by \citet{2011ApJ...727...49L}. The
far wing magnetograms are useful to detect pixels with strong
downflows but do not yield the magnitude of the velocities present in
those locations. In order to determine the velocities, the observed
Stokes profiles were subject to an inversion using the SIR code
\citep[Stokes Inversion based on Response Functions;][]{1992ApJ...398..375R}.  
Two sets of inversions were carried out. In the first run, a single
magnetic component was assumed in each pixel with the vector magnetic
field (field strength, inclination and azimuth) and LOS velocity
remaining constant with height. The resulting LOS velocity map is
shown in Fig.~2 of \citet{2011ApJ...727...49L}. Pixels exhibiting velocities
greater than 2~km~s$^{-1}$ were then selected and subject to a second
set of inversions in which two magnetic components were assumed to
co-exist in a single resolution element, both of which having
height-independent physical parameters. The one and two component SIR
inversions also retrieved height-independent micro- and
macro-turbulent velocities as well as the fraction of stray light in
each pixel. The left panel of Fig.~\ref{velo} shows the LOS velocity
combining the two inversion sets, while the fill fraction of the fast
component is depicted on the right.

\begin{figure}[t] 
\centering
\vspace*{-.5em}
\includegraphics[width=6.5cm,angle=90,bb=-11 -22 621 842]{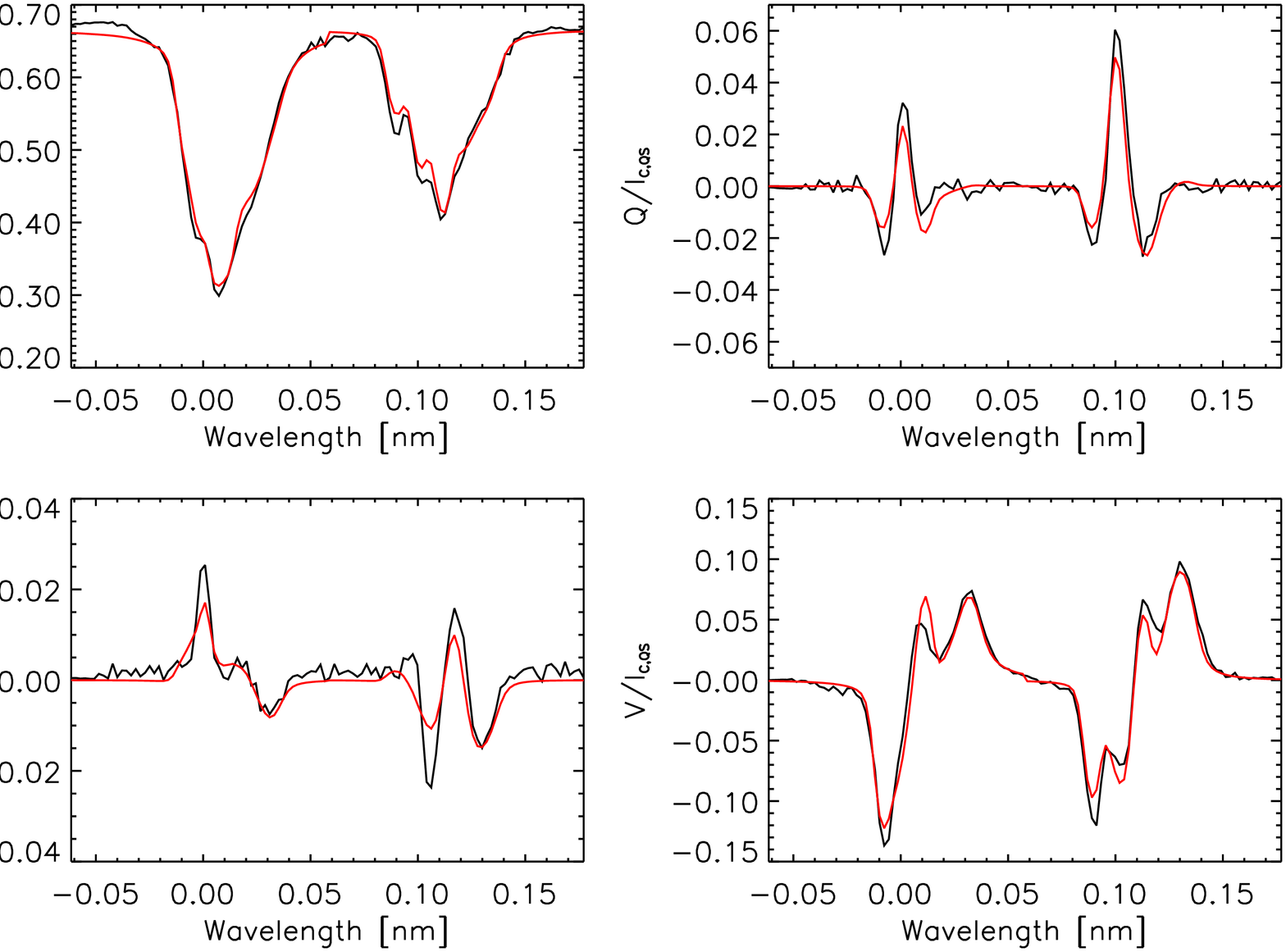} 
\includegraphics[width=6.5cm,angle=90,bb=-11 -15 621 850]{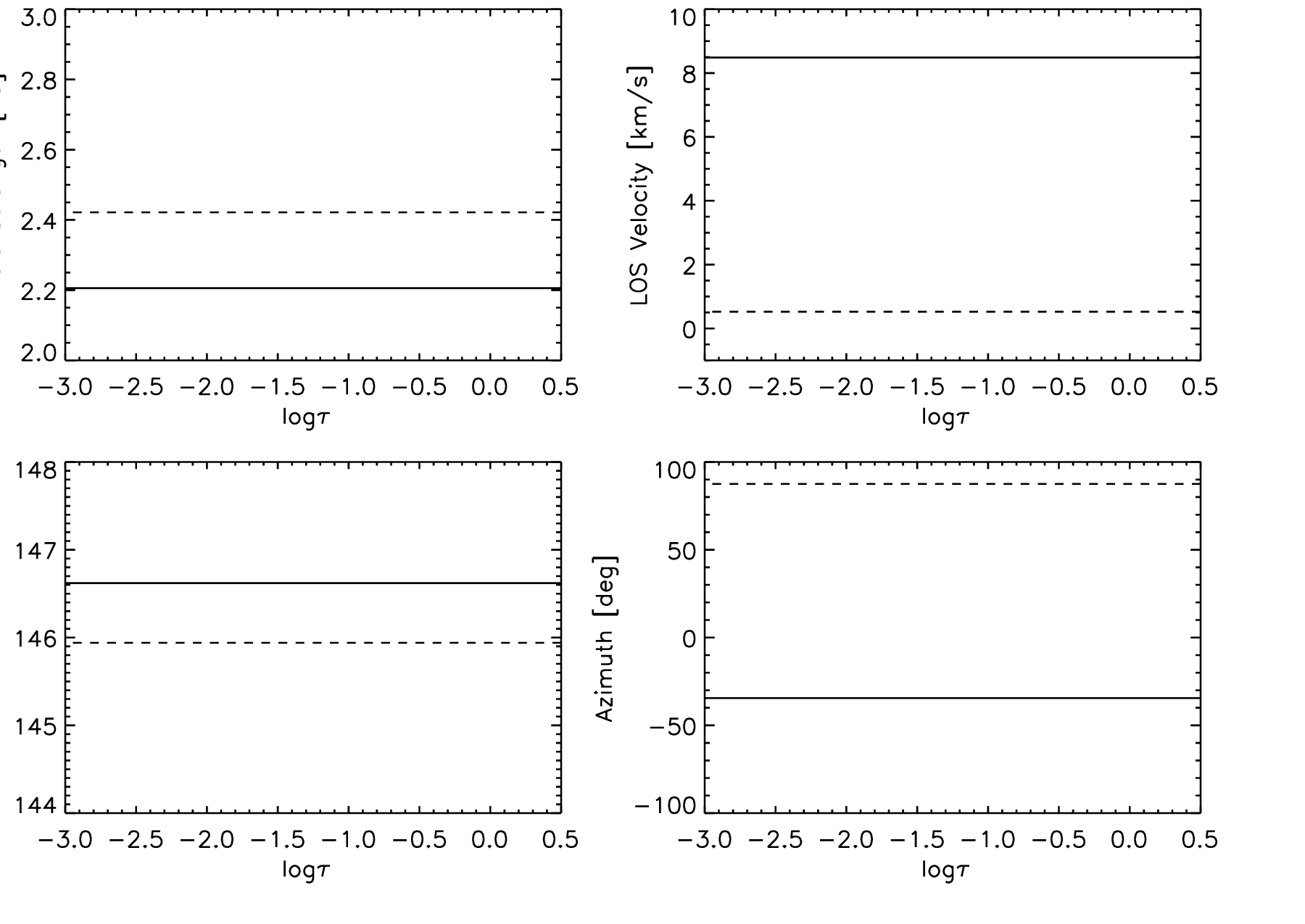}
\vspace*{-.5em} 
\caption{{\em First and second rows:} Observed (black) and best-fit (red) 
Stokes profiles of the pixel marked in Fig.~\ref{velo} using a
two-component model.  {\em Third and fourth rows:} Height-independent
stratifications of the physical parameters corresponding to the two
components (solid and dashed lines, respectively).}
\label{stokes} 
\end{figure} 
 
Figure~\ref{velo} reveals the existence of supersonic downflows of up
to $\sim$8~km~s$^{-1}$ in AR 10923 located nearly $3\arcsec$ from the
umbra-penumbra boundary.  To the best of our knowledge, these are the
largest downflows ever detected in the inner penumbra close to the
umbral boundary. The right panel of the figure indicates that a large
fraction of the resolution element is dominated by the stronger
downflowing component. We estimate the typical size of the downflowing
patches to be $\sim$1.6 arcsec$^2$. The strong downflowing zones are
surrounded by upflows measuring 2 km~s$^{-1}$ which can be identified
with the Evershed flow.

The Stokes $V$ profiles emerging from the downflowing regions exhibit
a satellite in the red lobe while the Stokes $I$ profiles have highly
inclined red wings, as illustrated in Fig.~\ref{stokes} for the pixel
marked with a cross in Figure~\ref{velo}. The above spectral
characteristics are reproduced satisfactorily by the two-component
inversions. The two magnetic components could either reside
side-by-side in the same resolution element or could be stacked one on
top of the other.  While the exact configuration remains uncertain,
supersonic velocities exist in the presence of very strong magnetic
fields as indicated by the third row of Figure~\ref{stokes}. In
addition, the polarity of the strong downflowing component is the same
as that of the parent sunspot, which rules out the possibility of them
being Evershed flows returning to the solar surface. The strong fields
exceeding 2~kG should also inhibit convection, suggesting that the
downflows are likely to be caused by an alternative mechanism.

\subsection{Photospheric and Chromospheric Brightenings} 
\label{activity} 
We now turn our attention to the photospheric and chromospheric
brightenings associated with the supersonic downflows. These
brightness enhancements are observed to lie in close proximity to the
downflowing patches and can have intensities comparable to the quiet
Sun. The penumbral filament near one of the downflowing regions has an
intensity of 0.9$I_{\textrm{\tiny{QS}}}$ in the continuum at 630~nm
(see inset of Figure~\ref{cont_image}). Higher up in the chromosphere,
these brightenings are $\sim$77\% more intense than the penumbral
microjets (MJs), which appear to be in the decay phase
\citep{2008ApJ...676.1356R}. While the brightness enhancements observed 
near the supersonic downflows appear as isolated blobs on the filaments, MJs
are oriented nearly perpendicular to the filament (see Fig.~4 of
\citealt{2008ApJ...676.1356R}).
 
\begin{figure}[t] 
\vspace{-15pt} 
\centerline{ 
\includegraphics[width=0.67\textwidth,angle=0]{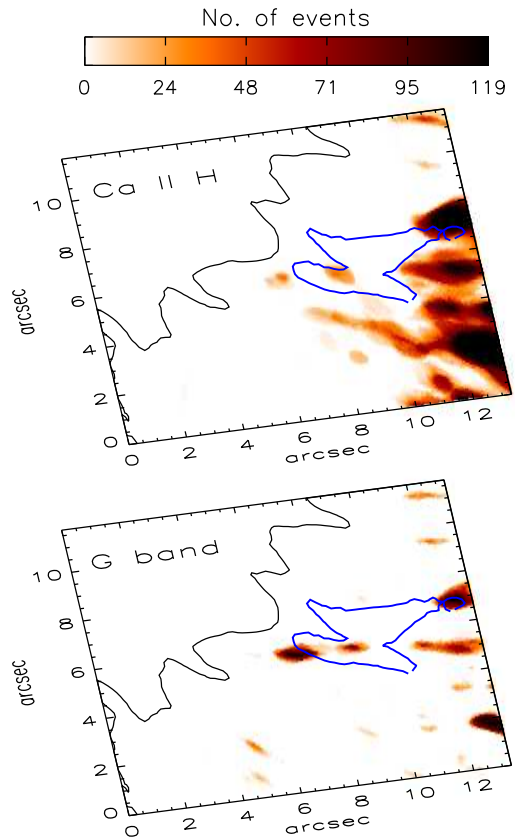} 
}
\vspace{-15pt} 
\caption{G-band ({\it{bottom}}) and Ca ({\it{top}}) event maps depicting 
locations with intensity close to the quiet Sun photosphere. Blue
contours of LOS velocity greater than 2 km~s$^{-1}$ have been overlaid
on the filtergrams. The black contour corresponds to the continuum
intensity at 630 nm. The images have been scaled as shown by the
horizontal color bar.}
\label{bright} 
\end{figure}

To determine if the proximity of the enhancements to the downflows
endures with time, event maps were constructed in the photosphere as
well as the chromosphere using the G-band and Ca filtergram time
sequence.  The procedure has been described in \citet{2011ApJ...727...49L} with
threshold values of 0.85 and 0.9 being employed for the photosphere
and chromosphere respectively.  The resulting event maps are shown in
Figure~\ref{bright}.
 
At both heights we find a large number of events concentrated near the
downflows resembling blobs that were seen in the individual
filtergrams.  While the brightenings in the photosphere persist for
the entire 1 hour sequence, in the chromosphere they last for only
about one third of the time. More importantly, the enhancements appear
very similar in shape and are nearly co-spatial.  The use of a large
threshold in the chromospheric event map removes all signatures of
the relatively weaker and transient MJs.
 
\section{Discussion} 
\label{orient} 
The downflows that are associated with the EF can sometimes be supersonic  
in the outer penumbra \citep{2001ApJ...549L.139D,2004A&A...427..319B} or even beyond the  
sunspot boundary \citep{2009ApJ...701L..79M}. Such a configuration represents mass flux 
returning to the photosphere and has a polarity opposite to that of the sunspot. 
The supersonic downflows we have observed have the same polarity as the parent sunspot 
and so they cannot be related to the Evershed downflows. One could assume that  
these strong downflows are the photospheric counterpart of some kind of inverse  
Evershed flow seen in the chromosphere. However, it is not clear how  
such a chromospheric phenomenon could produce supersonic downflows close to the 
umbra-penumbra boundary in the photosphere.

\begin{figure}[t] 
\centering
\vspace{-2pt} 
\hspace{20pt} 
\includegraphics[width=0.55\textwidth,angle=90]{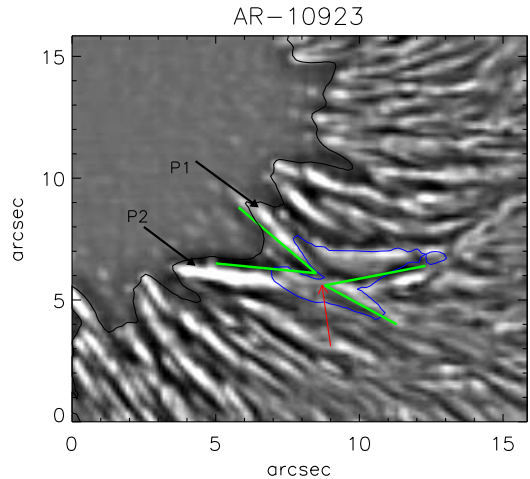} 
\vspace{-25pt} 
\caption{Continuum image that has been unsharp masked using a 
$3 \times 3$ pixel boxcar. The red arrow indicates the location where
the filaments P1 and P2 appear to intersect each other. The solid
green lines refer to the bisecting angles between the filaments P1 and
P2. The blue contour has been drawn for LOS velocities greater than 2
km~s$^{-1}$.}
\label{unsharp} 
\end{figure}

The orientation of the filaments P1 and P2, bifurcating at the strong
downflowing patch (Fig.~\ref{unsharp}), resembles the
post-reconnection configuration illustrated in Fig.~5c of
\citet{2008ApJ...686.1404R}, suggesting that the origin of the downflows is
the slingshot effect associated with the reconnection of the
filaments. The bisecting angles shown by the solid green lines were
estimated to be $51\deg$ and $46\deg$.
 
According to \citet{2008ApJ...686.1404R}, the unwinding of filaments in a  
cork screw fashion can lead to reconnection, transient brightenings  
and twists in the penumbral filaments. This model was proposed as a  
possible mechanism for producing penumbral MJs. \citet{2010ApJ...715L..40M} investigated 
the above scenario using numerical simulations and concluded that  
MJs occur in the intermediate region between nearly horizontal flux tubes and the 
relatively vertical background field of the penumbra. In this model, 
only parts and not the entire penumbral filament participate in the reconnection process. 
 
Slingshot reconnection may be a possible mechanism for producing the supersonic  
downflows and the photospheric as well as chromospheric brightenings, 
although there is no strict one-to-one correspondence between the two phenomena. 
The above process has to be different from the one producing MJs since 
their intensities and lifetimes are much smaller than the 
events described in this work.  
 
\section{Summary} 
\label{summary} 
High resolution spectro-polarimetric observations of NOAA AR 10923
taken with {\em Hinode} reveals supersonic downflows near the
umbra-penumbra boundary. The downflows are observed in large patches
having an area of 1.6 arcsec$^2$. Using a two-component model
atmosphere the SIR code retrieves supersonic values of
8~km~s$^{-1}$. These are the largest velocities ever detected at those
locations in a sunspot. Frequent occurrences of strong downflows at
the border of umbrae without penumbrae have been reported by
\citet{2008ApJ...680.1467S}. 
 
The strong velocities are associated with 2 kG magnetic fields which
have the same polarity as the sunspot. This would imply that the
downflows are not related to the Evershed flow, although the latter
could be separately present at those locations. We find intense and
long lived chromospheric brightenings near the strong photospheric
downflows extending over an area of 1--2 arcsec$^2$. Furthermore,
photospheric brightenings nearly as intense as the quiet Sun are also
present in the downflowing regions or close to them. These downflows
may have lifetimes of up to 14 hours (or more) as found from several
consecutive {\em Hinode}/SP scans
\citep{2011ApJ...727...49L}.
 
The penumbral filaments in the vicinity of the strong downflows appear
to be twisted in the manner described by \citet{2008ApJ...676.1356R} that
would arise from a reconnection process. Such a process could produce
the transient penumbral microjets. These events are ubiquitous in the
penumbra and the photospheric downflows associated with them are
typically 1 km~s$^{-1}$, as reported by \citet{2010A&A...524A..20K}. The
chromospheric brightenings described in Sect.~\ref{activity} however,
are stronger, bigger, and longer-lived than the microjets. The
supersonic downflows that we have observed are an entirely new
phenomenon. They are possibly driven by dynamic and very energetic
processes occurring in the inner penumbra which produce highly intense
and long duration brightenings in the photosphere as well as in the
chromosphere. A suitable theory is yet to be formulated for explaining
these events and will be a challenge for future numerical models of
sunspots.
 
\acknowledgements 
We sincerely thank the {\em Hinode} team for providing the high
resolution data. {\em Hinode} is a Japanese mission developed and
launched by ISAS/JAXA, with NAOJ as domestic partner and NASA and STFC
(UK) as international partners. It is operated by these agencies in
co-operation with ESA and NSC (Norway). We thank the organizers for a
wonderful meeting and for their warm hospitality.
 
\bibliography{louis} 
 
\end{document}